\documentclass[12pt,preprint]{aastex}

\newcommand{\teff}{\ensuremath{T_{\rm eff}}}

\newcommand{\rsun}{\ensuremath{R_\sun}}
\newcommand{\msun}{\ensuremath{M_\sun}}
\newcommand{\lsun}{\ensuremath{L_\sun}}

\newcommand{\rstar}{\ensuremath{R_\star}}

\newcommand{\kA}{KOI-74}
\newcommand{\kAb}{KOI-74b}
\newcommand{\kAsmass}{\ensuremath{2.22^{+0.10}_{-0.14}}} 
\newcommand{\kAsrad}{\ensuremath{1.899^{+0.043}_{-0.051}}} 
\newcommand{\kApmass}{\ensuremath{0.111^{+0.034}_{-0.038}}} 
\newcommand{\kAprad}{\ensuremath{0.0393^{+0.0009}_{-0.0013}}} 
\newcommand{\kAper}{\ensuremath{5.188754\pm0.000083}} 
\newcommand{\kAinc}{\ensuremath{88.8\pm0.5}} 
\newcommand{\kAepoch}{\ensuremath{2454958.87995\pm0.00036}} 
\newcommand{\kAeclip}{\ensuremath{1238.2\pm6.6}} 

\newcommand{\kAteffs}{\ensuremath{9400\pm150}}
\newcommand{\kAteffp}{\ensuremath{12289\pm340}}
\newcommand{\kALs}{\ensuremath{25.6\pm2.4}}
\newcommand{\kALp}{\ensuremath{0.0317\pm0.0060}}

\newcommand{\kB}{KOI-81}
\newcommand{\kBb}{KOI-81b}
\newcommand{\kBsmass}{\ensuremath{2.71^{+0.19}_{-0.11}}} 
\newcommand{\kBsrad}{\ensuremath{2.93\pm0.14}} 
\newcommand{\kBpmass}{\ensuremath{0.212\pm0.031}} 
\newcommand{\kBprad}{\ensuremath{0.115\pm0.006}} 
\newcommand{\kBper}{\ensuremath{23.8776\pm0.0020}} 
\newcommand{\kBinc}{\ensuremath{88.2\pm0.3}} 
\newcommand{\kBepoch}{\ensuremath{2454976.06981\pm0.00095}} 
\newcommand{\kBeclip}{\ensuremath{4977\pm16}} 

\newcommand{\kBteffs}{\ensuremath{10000\pm150}}
\newcommand{\kBteffp}{\ensuremath{13430\pm490}}
\newcommand{\kBLs}{\ensuremath{77.3\pm9.6}}
\newcommand{\kBLp}{\ensuremath{0.387\pm0.096}}

\shorttitle{Hot Compact Objects}
\shortauthors{Rowe et al.}
\slugcomment{Version 2.5 --- Janurary 19, 2009}

\begin{document}

\title{{\it Kepler} Observations of Transiting Hot Compact Objects}

\author{Jason~F. Rowe,\altaffilmark{1,2}
William~J. Borucki,\altaffilmark{2}
David~Koch,\altaffilmark{2}
Steve~B.~Howell,\altaffilmark{3}
Gibor~Basri,\altaffilmark{4}\\
Natalie~Batalha,\altaffilmark{5}
Timothy~M. Brown,\altaffilmark{6}
Douglas Caldwell,\altaffilmark{7}
William~D. Cochran,\altaffilmark{8}\\
Edward Dunham,\altaffilmark{9}
Andrea~K.~Dupree,\altaffilmark{10}
Jonathan~J.~Fortney,\altaffilmark{15}
Thomas~N.~Gautier~III,\altaffilmark{11}\\
Ronald~L.~Gilliland,\altaffilmark{12}
Jon~Jenkins,\altaffilmark{7}
David~W. Latham,\altaffilmark{10}
Jack~J.~Lissauer,\altaffilmark{2}
Geoff~Marcy,\altaffilmark{4}\\
David~G.~Monet,\altaffilmark{13}
Dimitar~Sasselov\altaffilmark{10} 
William~F.~Welsh,\altaffilmark{14}
}
\altaffiltext{1}{NASA Postdoctoral Program Fellow}
\altaffiltext{2}{NASA Ames Research Center, Moffett Field, CA 94035}
\altaffiltext{3}{National Optical Astronomy Observatory, Tucson, AZ 85719}
\altaffiltext{4}{University of California, Berkeley, Berkeley, CA 94720}
\altaffiltext{5}{San Jose State University, San Jose, CA 95192}
\altaffiltext{6}{Las Cumbres Observatory Global Telescope, Goleta, CA 93117}
\altaffiltext{7}{SETI Institute, Mountain View, CA 94043}
\altaffiltext{8}{University of Texas, Austin, TX 78712}
\altaffiltext{9}{Lowell Observatory, Flagstaff, AZ 86001}
\altaffiltext{10}{Harvard-Smithsonian Center for Astrophysics, Cambridge, MA 02138}
\altaffiltext{11}{Jet Propulsion Laboratory/California Institute of Technology, Pasadena, CA 91109}
\altaffiltext{12}{Space Telescope Science Institute, Baltimore, MD 21218}
\altaffiltext{13}{US Naval Observatory, Flagstaff Station, Flagstaff, AZ 86001}
\altaffiltext{14}{San Diego State University, San Diego, CA 92182}
\altaffiltext{15}{University of California, Santa Cruz, CA 95064}


\begin{abstract}
{\it Kepler} photometry has revealed two unusual transiting companions orbiting an early A-star and a late B-star.  In both cases the occultation of the companion is deeper than the transit. The occultation and transit with follow-up optical spectroscopy reveal a 9400 K early A-star, \kA\ (KIC~6889235), with a companion in a 5.2 day orbit with a radius of 0.08 \rsun\ and a 10000 K late B-star \kB\ (KIC~8823868) that has a companion in a 24 day orbit with a radius of 0.2 \rsun.  We infer a temperature of 12250 K for \kAb\ and 13500 K for \kBb.


We present 43 days of high duty cycle, 30 minute cadence photometry, with models demonstrating the intriguing properties of these object, and speculate on their nature.

\end{abstract}

\keywords{stars: individual (\kA,
KIC~6889235, \kB, KIC~8823868) --- Facilities: \facility{The Kepler Mission}.}

\section{Introduction}\label{intro}

The {\it Kepler} mission was designed to find transiting Earth-like planets in the habitable zones of other stars \citep{bor10}.  To accomplish this feat, {\it Kepler} was launched into a heliocentric, Earth-trailing orbit allowing stars to be continuously monitored for the lifetime of the mission.  A natural product of the mission is that any object that is Earth-sized or larger that transits its host star can be detected.  This includes stellar companions such as M-dwarfs and non-stellar companions such as brown dwarfs (BDs) and white dwarfs (WDs).


Observations of a stellar system with a smaller companion that orbits along our line-of-sight will be seen to move in front (transit) and behind the host star (occultation).  The transit lightcurve will have a curved "U" shape due to limb-darkening of the stellar surface.  The occultation will have a flat bottom and sharp egress and ingress as the flux contribution from the companion is completely obsured.  From the shape of the stellar transit the ratio of the companion and stellar radii and the scaled semi-major axis ($a/R_{\star}$) can be determined.  The term $a/R_{\star}$ is related to the mean stellar density. The depth of the occultation gives the ratio of the surface brightness of the companion and stellar host integrated over the observed bandpass.

We present 43 days of high duty cycle, ultra precise photometry of two transiting objects which have radii similar to Jupiter and effective temperatures $>$ 10000 K.  \kAb\ may have properties similar a low-mass white dwarfs, in that it is compact and hot and its radius combined with a rough estimate of its mass is consistent with an internal structure dominated by electron degeneracy. \kBb\ has a radius consistent with either a late M-dwarf, brown dwarf or Jupiter-sized planet, but this picture is inconsistent with its observed temperature.  

\section{{\it Kepler} Photometry and Transit Fits}\label{kepphot}

\begin{figure}
\begin{center}
\includegraphics[scale=0.9]{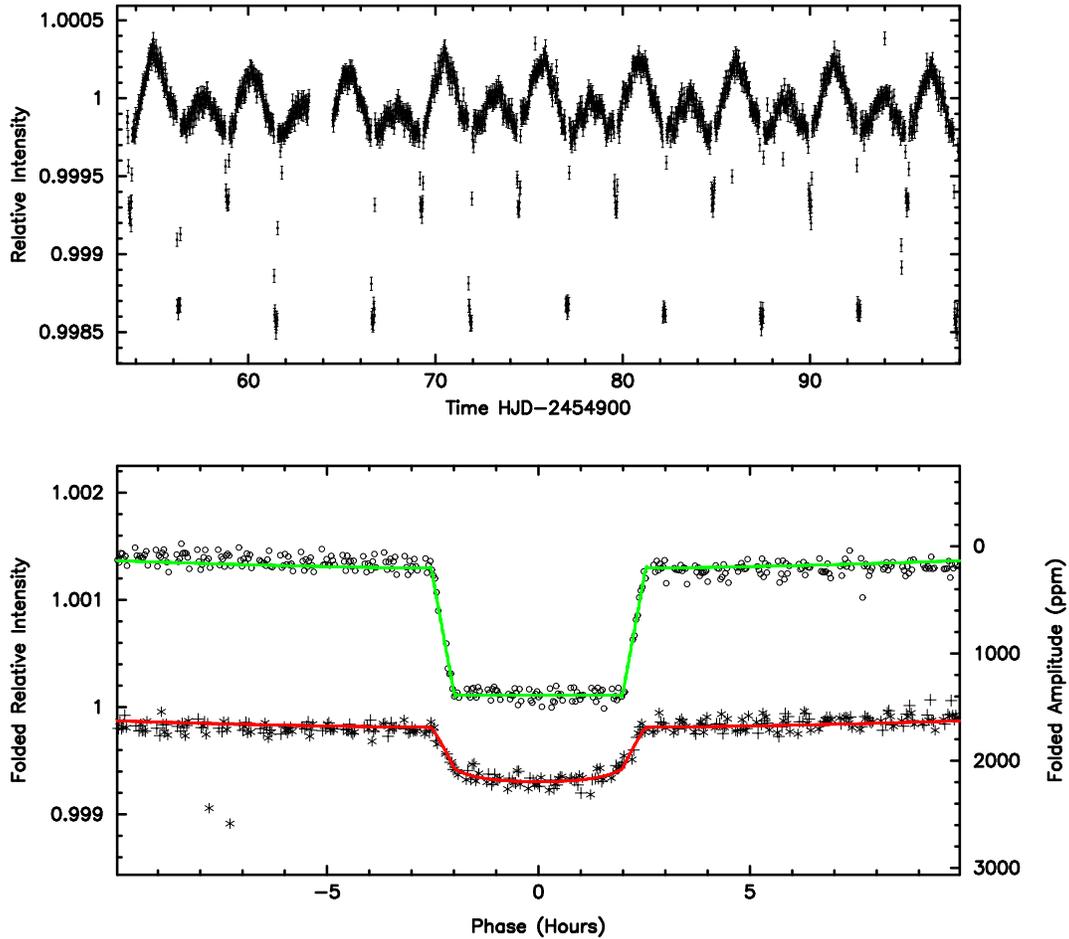}
\end{center}
\caption[KOI74]{The phased lightcurve of \kA\ containing 8 transits and 9 eclipses observed by the {\it Kepler} photometry between 2 May 2009 and 15 June 2009.  The upper panel shows the full 43-day time series after detrending.  The bottom panel shows the lightcurve folded with the orbital period.  The lower curve shows the primary eclipse, with the fitted transit model overplotted in red and corresponding scale to the left. The upper curve covers the expected time of secondary eclipse, with the fitted model overplotted in green with corresponding scale found to the right.  Odd and even transits are marked with stars and pluses, respectively, and measurements near the eclipse are shown with circles.}
\label{fig:koi74}
\end{figure}

\begin{figure}
\begin{center}
\includegraphics[scale=0.9]{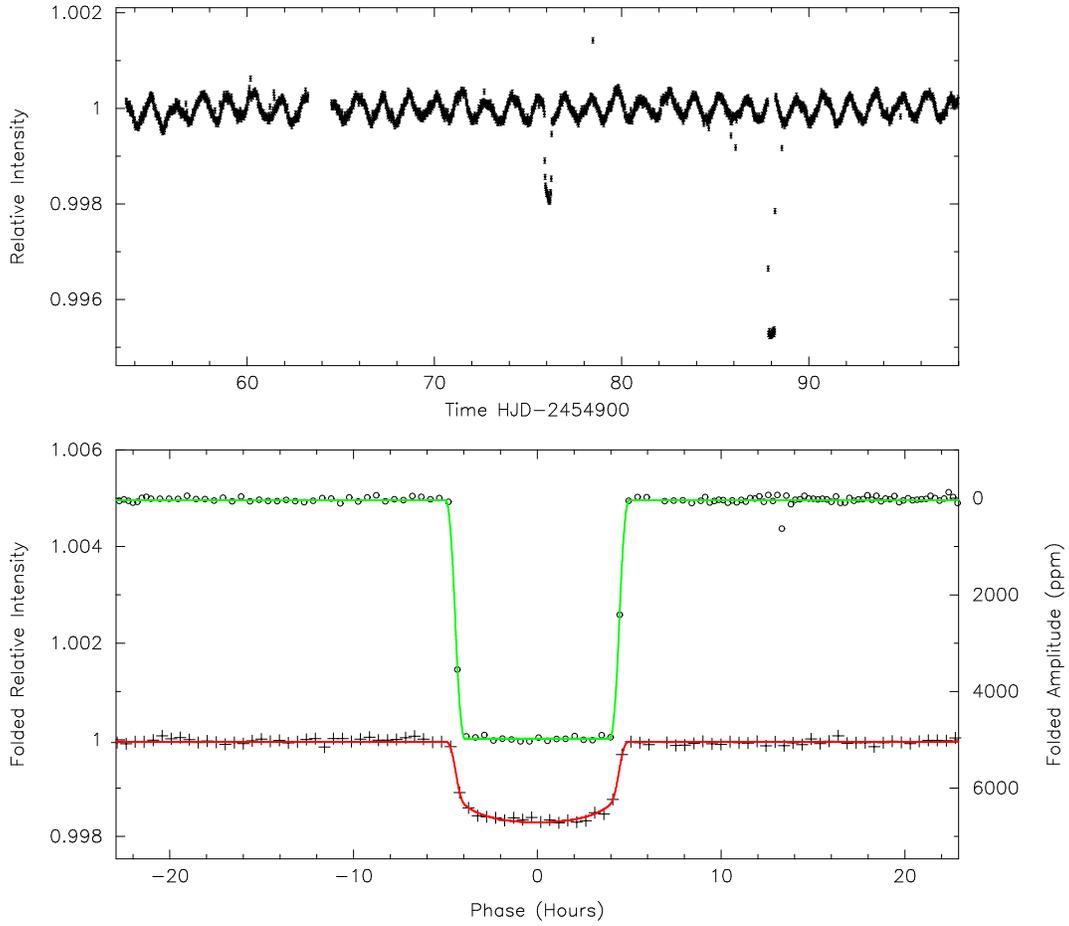}
\end{center}
\caption[KOI81]{The phased lightcurve of \kB\ containing a single transit and eclipse observed by the {\it Kepler} photometry between 2 May 2009 and 15 June 2009.  The upper panel shows the full 43-day time series after detrending.  The bottom panel shows the transit and eclipse after removal of periodicity seen 0.72 c/d.  See Figure \ref{fig:koi74} for further details.}
\label{fig:koi81}
\end{figure}

The top panels of Figures \ref{fig:koi74} and \ref{fig:koi81} show 43 days of photometry from the {\it Kepler} instrument with one-sigma error bars.  The observations have a cadence of 30 minutes.  The gap seen near HJD 2454965 is due to the transmission of data to the Earth and marks the gap between {\it Q0} and {\it Q1} observations \citep{koc10}.  The lightcurve for \kB\ shows periodicity at 0.72 d.  This signal was removed prior to the transit fits by masking off the transit and eclipsing and fitting a sinusoidal curve to the lightcurve.  

The bottom panels of Figures \ref{fig:koi74} and \ref{fig:koi81} show the observations phased with the orbital period. The bottom curve and data are centred on the transit, whereas the upper curve and data are centred on the occultation.  The points near transit are shown with stars and pluses where the different symbols indicate odd and even transit respectively.  Measurements taken during the occultation are shown with open circles.

The data were filtered with a running median boxcar with a width of 10 days.  Transit models are computed using the analytic formalue of \citet{man02} with non-linear limb darkening.  The data is fit for the center of transit time, period, impact parameter ($b$), the scaled planetary radius (R/R$_{\star})$ and $\zeta$/R$_{\star}$.  The last term, $\zeta$/R$_{\star}$ is related to the transit duration ($T_d=2(\zeta/\rstar)^{-1}$) and the mean stellar density \citep{pal09}.  The transit models are shown in Figures \ref{fig:koi74} and \ref{fig:koi81} by the red (lower) line.  The occultation is modeled by assuming no limb darkening for the occulted object and shown by the green (upper) line.  The transit model assumes that the companion does not emit.  While not true, this is a good approximation as the companions, \kAb\ and \kBb\ are 800 and 200 times less luminous that the host stars thus the dilution only alters the companion radius at the 1\% level.  We also assumed a circular orbit in our model fits.  

Optical spectroscopy obtained at the Kitt Peak 2.1-m was used to estimate \kA\ as spectral type A1V with \teff= \kAteffs\ K and \kB\ as B9-A0V with \teff=\kBteffs K.  With the stellar density from the transit light curve and \teff\ the stellar mass and radius where estimated using the $\rho_{\star}$ method as described in \citet{bor10b}.  This allows us to model the companion radii, orbital period, inclination angle and eclipse depth which we list in Table 1 as described in \citet{row10}.

\begin{figure}
\begin{center}
\includegraphics[scale=0.9]{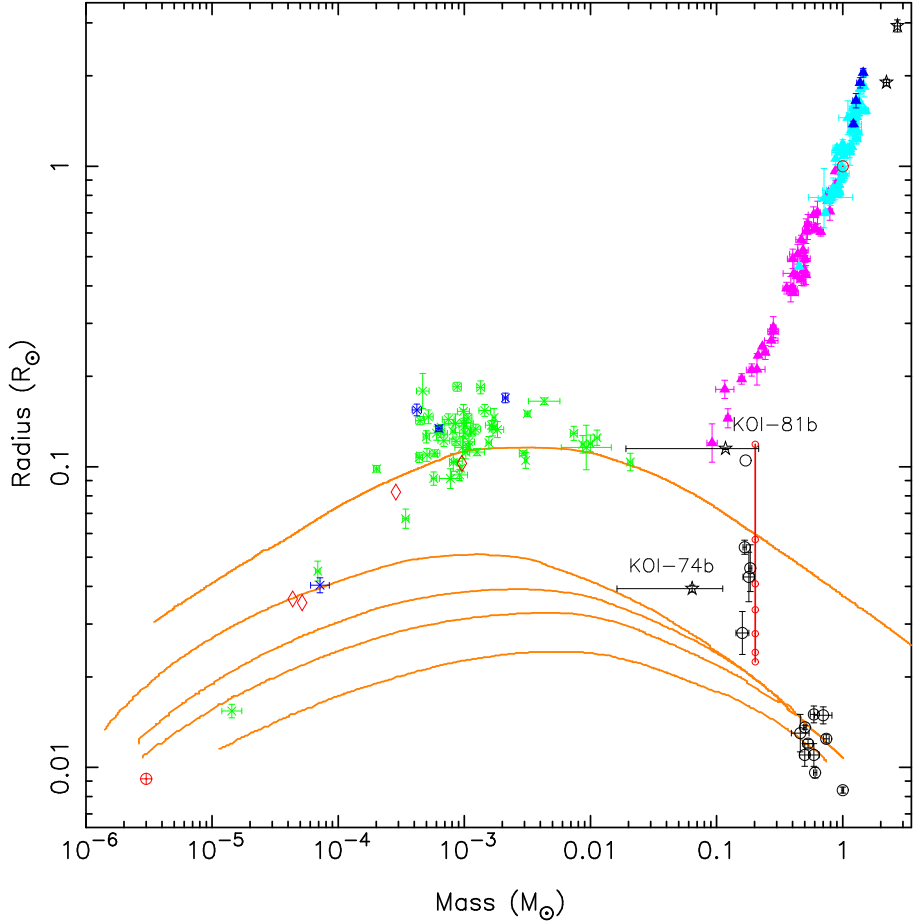}
\end{center}
\caption[Mass Radius Diagram]{Location of \kAb\ and \kBb\ on the mass-radius diagram.  Stars with transiting exoplanets are show in cyan, transiting exoplanets with well defined masses and radii are shown in green. {\it Kepler} transiting planets are shown in blue.  The positions of the Earth, Uranus, Neptune, Saturn, Jupiter and the Sun are also indicated by red diamonds.  Low-mass stars from \citep{lop07} are shown in magenta.  White dwarfs as observed by Hipparcos \citep{pro98} and a sample of extremely-light WDs are shown with black \textbf{o}'s. The open circle near \kBb\ is the millisecond pulsar companion discovered by \citet{edm01}. The orange lines, from top to bottom, show zero temperature models of
\citet{zp69} with compositions of H, He, Mg, C and Fe. The red vertical line is the cooling curve for a 0.2026 \msun\ He-WD from \citep{pan07}. \kA\ and \kB\ are shown as stars in the upper right portion of the diagram and the positions of \kAb\ and \kBb\ are labeled.}
\label{fig:massrad}
\end{figure}

\begin{table}\label{ta:pars}
\begin{center}
\begin{tabular}{lcc}
\hline \hline \kA &  & \\
  & Star & Companion \\
\hline
Radius   & \kAsrad\ \rsun & \kAprad\ \rsun \\
Mass     & \kAsmass\ \msun & 0.02-0.11 \msun \\
Luminosity & \kALs\ \lsun & \kALp\ \lsun \\
\teff & \kAteffs\ K & \kAteffp\ K \\
Epoch & & \kAepoch\ HJD \\
Period & & \kAper\ days \\
{\it i} & & \kAinc\ deg \\
Eclipse Depth & & \kAeclip\ ppm \\
\hline \kB &  & \\
  & Star & Companion \\
\hline
Radius   & \kBsrad\ \rsun & \kBprad\ \rsun \\
Mass     & \kBsmass\ \msun & 0.02-0.2 \msun \\
Luminosity & \kBLs\ \lsun & \kBLp\ \lsun \\
\teff & \kBteffs\ K & \kBteffp\ K \\
Epoch & & \kBepoch\ HJD \\
Period & & \kBper\ days \\
{\it i} & & \kBinc\ deg \\
Eclipse Depth & & \kBeclip\ ppm \\
\hline\\
\end{tabular}
\caption[System Parameters for KOI 74]{Modeled parameters for \kA\ and \kB. The stellar radius, mass and luminosity are determined from the $\rho_{\star}$ method.  The stellar temperatures are spectroscopically determined.  The radius of the companion, Period, {\it i}, Epoch and Eclipse Depth are based on fits to the transit and eclipse photometry. $M_p$ is estimated from the amplitude of variations phase locked to the orbital period. The companion luminosities and temperatures are estimated assuming a bolometric eclipse depth and blackbody behaviour.}
\end{center}
\end{table}

\subsection{Temperatures of the Companions}

The depth of the eclipse is deeper than the transit for both \kA\ and \kB, indicating the companions have a larger surface brightness as compared to the stellar host.  The spectroscopically determined effective temperature and the transit determined stellar density with the $\rho_{\star}$ method provide luminosity estimates of the stars. If we assume that the eclipse depths are bolometric, than the depth of the eclipse gives the luminosity ratio of the star and companion.  The estimated luminosities of the stars and companions are listed in Table 1.  Assuming the companions act as blackbodies and using the radii determined from the transit fit we find temperatures of \kAteffp\ K and \kBteffp\ K for \kAb\ and \kBb.
 
\subsection{Mass Estimation}\label{massest}

\kA\ and \kB\ exhibit variations with a periodicity equal to one-half the orbital period and presenting amplitudes of $\sim$193 ppm and $\sim$50 ppm respectively.  The \kA\ lightcurve also exhibits variability with an amplitude of ~116 ppm with a period equal to the orbital period.  There are various ways to interpret these variations. We could be seeing spot activity induced on the stellar surface from a magnetic interaction between the star and companion.  This would be similar to activity seen in the planet hosting system HD179949 \citep{shk04} or $\tau$ Bootis \citep{wal08}.   The variations could also be phase locked due to tidal distortions of the host star.  In the later scenario we can attempt to estimate the mass of the companion.  

Ellipsoidal variations show two maxima and minima per orbital revolution.  This variability is due the changing observed stellar cross section, reflection effects due to emission and absorption from each component and gravity darkening.  \citet{mor85} found from a cosine expansion of the double wave characteristic of the variations that the amplitude of the light variations are given by $\Delta M = f(\tau,U_1)qR^3_{\star}sin^3(i)/a^3$, where, $q$ is the mass ratio of the companion and star and $\tau$ and $U_1$ express the gravity and limb-darkening terms that describe the surface profile of the distorted star with a radius of $R_{\star}$ with a companion in an orbit with inclination $i$ with a semi-major axis, $a$.  Adopting gravity and limb-darkening  co-efficients from Table II of \citet{bee89} we estimate the mass of \kAb\ $\sim 0.08$ \msun\ and \kBb\ $\sim 0.19$ \msun.  These relations work when the assumption that the tidally perturbed stellar surface remains in hydrostatic balance.  An important issue we address below. 

The ratio of the tidal acceleration to the star's surface gravity is given by,
\begin{equation}
\epsilon\equiv\frac{M_p}{M_{\star}} \left( \frac{R_{\star}}{a} \right) ^3,
\end{equation}
where $M_{\star}$ and $M_p$ are the mass of the star and companion, $R_{\star}$ is the stellar radius and $a$ is the semi-major axis.  This gives a rough measure of the amplitude of the ellipsoidal variability.  Assuming that the system is in tidal equilibrium, Equation 16 of \citet{pfa08} can be also be used to estimate the mass of the system. Adding this prescription to our transit model allows us to fit for the companion mass.  We find that \kAb\ has a mass of \kApmass\ \msun\ and  \kBb\ has a mass of \kBpmass \msun.

The assumption of tidal-equilibrium works well for short period binaries where both members have similar masses.  Tidal dissipation circularizes the orbit and synchronizes the stellar rotation period to the orbital period.  When the masses are not similar there is no reason to expect the rotation and orbital periods to synchronize, thus the assumption of tidal-equilibrium may be invalid. 

As shown by \citet{pfa08}, the tidal-equilibrium model works well for stars with deep convective atmospheres where flux pertrubations arise from changes in the local effective gravity.  This is not the case for stars such as \kA\ and \kB\ with radiative atmospheres.  Tidal forcing of radiative regions could produce significant deviations from hydrostatic balance as gravity waves can penetrate the interiors.  Simulations indicate that the observed oscillations can be greater than 10$\epsilon$ for stars with M$>$1.4\msun.  

If the companion is the remnant of a much larger star and has experienced an episode of mass loss, then the rotational and orbital periods may have already been synchronized.  In this case, the companions likely have significant mass.  We plan to obtain follow-up spectroscopic observations to search for radial velocity changes.  Such observations will place firm limits on the mass of the companions.  Until then, we adopt an uncertainty of 10$\epsilon$ in our estimate for the masses of the companions listed Table 1 to account for the possibility that the flux variations are produced by g-modes propagating to the surface of the star.   

\section{Discussion}

The temperatures of the companions are hot.  If these objects are planets then we can investigate their thermal properties when the only energy source is irradiation from the star.  Assuming a Bond Albedo of zero and complete redistribution of heat for reradiation we can calculate the equilibrium temperature ($T_{eq}$) assuming the companions are blackbodies.  The light curve does not show any indication of variability at the orbital period that might indicate a day-night temperature gradient for a tidally locked companion.  The calculation of $T_{eq}$ estimates the temperature of the planet assuming the only energy input is flux from the host star.  For \kAb\ this gives an equilibrium temperature of $2250 \pm 50$ K and $1715 \pm 50$ K for \kBb.  Comparing these values to the measured effective temperatures allows us to conclude that the objects require an additional energy source.   This could indicate prior evolution of the companions and we are watching the objects slowly cool.  

Figure \ref{fig:massrad} shows the positions of the host stars and companions relative to stars, planets and white dwarfs.  The green and cyan points are previously known transiting extrasolar planets\footnote{The Extrasolar Planets Encyclopedia: http://www.exoplanet.eu} and their host stars.  The dark blue points show the first five {\it Kepler} extrasolar planets \citep{bor10}.  The red diamonds indicate the Earth, Uranus, Neptune, Saturn and Jupiter.  The magenta points mark low-mass stars from \citet{lop07}.  The black circles clustered near 1 \msun\ are WDs observed by Hipparcos \citep{pro98} and those near 0.2 \msun\ are a sample of extremely low-mass white dwarfs \citep{edm01,bas06,lie04,kaw09,ker96} that have degenerate companions. The orange lines, from top to bottom, show zero temperature white-dwarf models of \citet{zp69} with compositions of H, He, Mg, C and Fe. The red vertical line is the cooling curve for a 0.2026 \msun\ He-WD from \citep{pan07}. The red vertical line shows the cooling model for a He-WD with a mass of 0.2026 \msun\ formed through mass-transfer \citep{pan07}.  The stars \kA\ and \kB\ are denoted in the upper right portion of the diagram by black stars and the positions of the companion \kAb\ and \kBb\ are labeled.  


The position of \kBb\ in the mass-radius diagram is similar to low mass stars at the hydrogen burning limit but its inferred temperature is much higher.  It has similar physical properties to the cooling He-WD discovered in the globular cluster 47 Tuc by \citet{edm01}.  Stellar modeling from the $\rho_{\star}$ \citep{bor10b} method puts the age of the host star at $340\pm67$ Myr.  The observed mass and radius of the companion estimate $\sim$155 Myr since its creation from the cooling curves of \citet{pan07}.  \kAb\ and \kBb\ orbit at distances of 16 \rsun\ and 49 \rsun, respectively.  When the hot stars begin shell burning of hydrogen in their interiors, their radii will dramatically increase and will likely result in mass transfer to the companions. That these objects are both found around hot stars is intriguing. The companions are inundated with UV radiation from the host star and are likely suffering from substantial evapouration and ionization.  

The lightcurve for \kB\ shows significant power near 0.72 c/d. These may be stellar pulsations induced by tidal interactions with the companion similar to the activity has been seen in the F-type binary star system HD~209295 \citep{han02}.  Observations of stellar oscillations offer the opportunity to refine the stellar parameters through asteroseismology.

\kA\ and \kB\ are interesting astrophysical systems.  Confirmation of the companion masses with spectroscopic radial velocities will help confirm the physical properties of these objects.  Follow-up observations are planned as well as continued observations with the {\it Kepler} instrument to help unravel their nature.

\acknowledgments

Funding for this Discovery mission is provided by NASA's Science Mission Directorate.  We are indebted to the entire {\it Kepler} Team for all the hard work and dedication have made such discoveries possible.

\end{document}